\documentclass[twocolumn,aps,prd,preprintnumbers,superscriptaddress,nofootinbib,amsmath,amssymb,floats,floatfix,notitlepage,longbibliography]{revtex4}

\usepackage{orcidlink}
\usepackage{lipsum}
\usepackage{graphicx}
\usepackage{subfigure}
\usepackage{braket}
\usepackage[utf8]{inputenc}
\usepackage{sans}
\usepackage{float}
\usepackage{changes}
\usepackage{hyperref}
\hypersetup{colorlinks=true,linkcolor=blue,urlcolor=blue,citecolor=blue}
\usepackage[toc,page]{appendix}
\usepackage[normalem]{ulem}
\usepackage{adjustbox}
\usepackage{latexsym}
\usepackage{amsmath}
\usepackage{amssymb}
\usepackage{amsfonts}
\usepackage{dcolumn}
\usepackage{bm}
\usepackage{tikz}
\usepackage{bigints}
\usepackage{array,tabularx,multirow,booktabs}
\usepackage[tracking=true]{microtype}
\SetTracking{}{500}
\SetTracking{encoding={*}, shape=sc}{40}
\UseRawInputEncoding 
\allowdisplaybreaks

\begin{document} \sloppy

\title{Infrared Extended Uncertainty Principle Corrections and Quintessence-Induced Topology of Reissner-Nordstr\"om AdS Black Holes
}

\author{Y. Sekhmani\orcidlink{0000-0001-7448-4579}}
\email[Email: ]{sekhmaniyassine@gmail.com}
\affiliation{Center for Theoretical Physics, Khazar University, 41 Mehseti Street, Baku, AZ1096, Azerbaijan.}
\affiliation{Centre for Research Impact \& Outcome, Chitkara University Institute of Engineering and Technology, Chitkara University, Rajpura, 140401, Punjab, India.}
\affiliation{Institute of Nuclear Physics, Ibragimova, 1, 050032 Almaty, Kazakhstan}
\affiliation{Department of Mathematical and Physical Sciences, College of Arts and Sciences, University of Nizwa, P.O. Box 33, Nizwa 616, Sultanate of Oman}

\author{G.~G.~Luciano\orcidlink{0000-0002-5129-848X}}\thanks{Corresponding author}
\email[Email: ]{giuseppegaetano.luciano@udl.cat }
\affiliation{Departamento de Qu\'{\i}mica, F\'{\i}sica y Ciencias Ambientales y del Suelo, Escuela Polit\'ecnica Superior -- Lleida, Universidad de Lleida, Av. Jaume II, 69, 25001 Lleida, Spain
}

\author{S. N. Gashti\orcidlink{0000-0001-7844-2640}}
\email[Email: ]
{sn.gashti@du.ac.ir; saeed.noorigashti70@gmail.com }
\affiliation{School of Physics, Damghan University, P. O. Box 3671641167, Damghan, Iran}

\author{A. Baruah, \orcidlink{0000-0001-6420-7666}}
\email[Email: ]{ anshuman.baruah@aus.ac.in}
\affiliation{Department of Physics, Assam University, Cachar - 788011, Assam, India}

\date{\today}

\begin{abstract}
We present a unified topological and geometric analysis of charged Anti-de Sitter (AdS) black holes immersed in a quintessence field, incorporating infrared gravitational corrections arising from the Extended Uncertainty Principle (EUP). The latter modifies the standard Heisenberg uncertainty relation by introducing a minimal momentum/maximal length scale, which effectively captures long-wavelength quantum gravitational effects relevant to black hole thermodynamics in curved spacetimes. We derive analytic expressions for the corrected Hawking temperature, entropy and heat capacity in terms of the EUP deformation parameter. Furthermore, the inclusion of quintessence, characterized by barotropic indices \(\omega_q = -\frac{2}{3}\) and \(\omega_q = -\frac{1}{3}\), modifies the black hole metric function. By studying the relaxation-time function $\tau(r_h)$, we identify a number of inflection points that depends sensitively on the equation of state parameter of quintessence, indicating a nontrivial impact of the latter on the black hole phase structure.
Applying Duan's topological current method to the off-shell free energy, we compute integer-valued winding numbers associated with each thermodynamic critical point. 
A parallel topological analysis of the photon sphere assigns charges \( \pm 1 \) to individual light rings, showing that quintessence effects can trigger the splitting or merging of photon spheres, while preserving the total exterior topological charge of $-1$.
\end{abstract}


\maketitle

\section{Introduction}
The thermodynamics of black holes has emerged as a cornerstone in the intersection of gravity, quantum theory and statistical physics. Seminal works by Bekenstein and Hawking first established that black holes possess an entropy proportional to their horizon area and emit thermal radiation with a temperature governed by their surface gravity~\cite{Bekenstein:1973ur,Hawking:1975vcx}. In asymptotically AdS spacetimes, these thermodynamic quantities exhibit behaviors reminiscent of classical thermodynamic systems, such as the Van der Waals fluid analogy observed in Reissner-Nordstr\"om AdS black holes~\cite{Kubiznak:2012wp,Chamblin:1999tk} (see also~\cite{Luciano:2023fyr,Luciano:2023bai} for applications in more general non‑extensive statistical frameworks).


Beyond the semiclassical approximation, modified gravity is expected to introduce corrections to the standard thermodynamic framework. The Extended Uncertainty Principle (EUP), which incorporates a minimum momentum or maximum length scale, modifies the canonical commutation relations and leads to significant infrared (IR) corrections in black hole thermodynamics~\cite{EUP1,EUP2,Mignemi:2009ji,EUP3,Mureika:2018fri,Arraut:2010gv}. In this framework, the Hawking temperature and entropy acquire deformation-dependent corrections governed by the EUP~\cite{Mignemi:2009ji}. These corrections are not merely formal; they potentially alter black hole phase transitions, horizon stability and observable quantities such as shadow radii and quasinormal modes~\cite{Nozari:2012nf,Mureika:2022rty}. As such, the study of EUP-deformed black holes provides a valuable phenomenological window into the interplay between IR gravity effects and horizon-scale physics, offering predictions that may be tested through astrophysical and gravitational wave data.

On the other hand, the presence of exotic fields, such as quintessence modeled as a scalar field with negative pressure and barotropic index \(\omega_q \in (-1,-1/3)\), has been extensively studied as an effective description of dark energy~\cite{Caldwell:1997ii}. The Kiselev solution describes black holes surrounded by such a field, leading to modifications in the spacetime geometry that significantly impact both the thermodynamic and causal structure of the black hole~\cite{Kiselev:2002dx}. In particular, different values of \(\omega_q\) induce distinct corrections to the metric function and result in qualitatively different phase structures~\cite{Zeng:2016aly,Shi:2019}. Moreover, weak gravitational lensing has been investigated in the background of a Kerr–Newman black hole surrounded by quintessential dark energy~\cite{Javed}.

Although previous studies have independently investigated the thermodynamic corrections induced by the EUP~\cite{EUP1,EUP2,EUP3} and the properties of Reissner-Nordström-AdS black holes in the presence of quintessence~\cite{Caldwell:1997ii,Kiselev:2002dx}, a comprehensive framework that simultaneously incorporates both effects and analyzes the resulting phase structure using topological methods has yet to receive adequate attention.
Such an investigation is particularly significant because both the EUP and quintessence arise from independent considerations in quantum gravity and cosmology, respectively, and each introduces profound modifications to black hole thermodynamics. Their combined effect could unveil  topological features that are not evident when considering either component in isolation. Therefore, a comprehensive analysis not only enriches our understanding of black hole phase transitions in modified gravity scenarios but also provides a fertile ground for testing the interplay between quantum-scale corrections and dark energy-like fields in a consistent thermodynamic framework.

To study black hole thermodynamics and phase transitions from a topological perspective, we adopt Duan's \(\phi\)-mapping theory, which translates the behavior of the thermodynamic scalar potential into a vector field defined on the \((T,r_h)\) plane. The zeros of this field correspond to thermodynamic critical points and are assigned integer winding numbers: \(+1\) for conventional phase transitions and \(-1\) for novel or inverse transitions. These topological charges remain invariant under smooth deformations of system parameters such as the deformation parameter \(\alpha\), the barotropic index \(\omega_q\) and the quintessence normalization constant \(\chi\)~\cite{Duan:1998jx,Wei:2022dzw,Wei:2023env}. This approach provides a robust topological classification scheme that complements and extends conventional thermodynamic analysis. In passing, we note that similar topological methods have been successfully applied in condensed matter systems, including the classification of interacting Majorana models and the identification of nonlocal order parameters in topological superconductors~\cite{Kita,Maiellaro:2022ake,Maiellaro:2022gmp}, as well as in topology-enhanced superconducting qubit networks~\cite{Settino}.

Compared to conventional approaches relying on the analysis of response functions (such as heat capacity divergences) or the behavior of thermodynamic potentials, Duan's topological framework offers several notable advantages. First, it provides a coordinate-invariant and metric-independent classification of critical points through topological invariants, making it particularly robust under reparametrizations or deformations of the thermodynamic phase space. Furthermore, the assignment of integer winding numbers to critical points captures not only their location but also their qualitative nature, distinguishing between conventional and inverse phase transitions in a mathematically rigorous manner. Finally, this method remains well-defined even in cases where standard thermodynamic quantities become non‑analytic or fail to exhibit clear divergences, thereby extending the reach of phase structure analysis into regimes inaccessible via conventional tools~\cite{Duan:1998jx,Wei:2022dzw,Wei:2023env}.

Topological methods have proven insightful not only in thermodynamic phase analysis but also in the study of the causal and optical structures of black holes \cite{P5}. In particular, the photon sphere, i.e., the region corresponding to unstable circular light orbits, exhibits a nontrivial topological structure. Each photon sphere is associated with a topological charge, characterized by an integer winding number defined in the impact-parameter space of null geodesics. These winding numbers encode the qualitative behavior of light propagation near the black hole and offer a topological classification of photon orbits. Notably, the sum of the winding numbers remains conserved under smooth deformations of the spacetime geometry, and the total exterior topological charge remains fixed at \(-1\), in accordance with the Weak Cosmic Censorship Conjecture~\cite{Wei:2020ght,NooriGashti:2024gnc}. This conservation law reflects the stability of the causal boundary and reinforces the idea that topological invariants can serve as powerful tools in probing the fundamental consistency of black hole spacetimes.

Starting from the above premises, in this work we present a systematic study of charged AdS black holes under simultaneous EUP and quintessence deformations, employing Duan's topological current method and geometric invariants of the photon sphere. Specifically:
\begin{itemize}
  \item We derive analytic, closed-form expressions for the EUP-corrected Hawking temperature, entropy and heat capacity in the presence of quintessence. Furthermore, by applying Duan's topological current method to the off-shell free energy, we compute integer-valued winding numbers for each thermodynamic critical point.
  \item We  perform a parallel topological analysis of the photon sphere, assigning \(+1\) and \(-1\) charges to individual light rings. In this context, we focus on the corrections induced by quintessence, which can significantly affect the effective potential for photons. We disregard the EUP-induced infrared corrections, as they are expected to primarily influence the asymptotic structure of spacetime and have minimal impact on the local geometry near the photon sphere. Interestingly, we show that quintessence can induce the splitting or merging of photon spheres, while preserving the total exterior topological charge of \(-1\).
\end{itemize}

Throughout this manuscript, we adopt natural units by setting \(\hbar = c = G = k_B = 1\).

{\section{Extended Uncertainty Principle and Quintessence: A Brief Overview}
Thermal radiation from black holes arises as a quantum phenomenon. To accurately capture the quantum behavior of the emitted particles, it is natural to expect that their position and momentum operators should satisfy the fundamental commutation relation $[\hat x,\hat p]=i$, which, in turn, gives the Heisenberg Uncertainty Principle (HUP)
\begin{equation}
\label{ineq}
  \Delta x\,\Delta p\ge\frac{1}{2}.
\end{equation}

However, in scenarios involving gravitational systems with large characteristic lengths - such as black holes in asymptotically AdS spacetimes or in the presence of dark energy components - it becomes relevant to consider potential infrared (IR) modifications to the uncertainty principle. Such modifications are motivated by various approaches to extended gravity, which suggest the existence of a minimal momentum or a maximal length scale. Within this context, the EUP offers a natural framework to incorporate IR corrections into quantum mechanics~\cite{EUP1,EUP2}. 

In the EUP framework, the HUP \eqref{ineq} is modified in the following form~\cite{EUP1,EUP2,Mignemi:2009ji,EUP3,Mureika:2018fri,Arraut:2010gv}:
\begin{equation}
\label{Ineq2}
  \Delta x\,\Delta p\ge\frac{1}{2}\left(1+\frac{\alpha}{L^2}\Delta x^2\right),
\end{equation}
where $\alpha$ is the deformation parameter and $L$ represents a characteristic length scale.  
In the case where \( \alpha/L^2 > 0 \) (corresponding to an AdS background), the above relation
indicates the presence of a lower bound on the momentum uncertainty. Conversely, when \( \alpha/L^2 < 0 \)  (associated with dS space), a maximal length scale naturally arises, identified with the horizon radius \( L \). Notably, the presence of a minimal momentum uncertainty in anti-de Sitter space can be linked to the existence of a lower mass limit for fields propagating in such a background~\cite{Breit}.
Clearly, setting $\alpha=0$ recovers the usual HUP~\eqref{ineq}.
In this work, we focus on the case \( \alpha > 0 \), which is particularly suited for studying the impact of long-wavelength quantum corrections on the thermodynamic topology of AdS black holes, where phase transitions and stability properties are known to be sensitive to such effects.

In passing, we note that Eq.~\eqref{Ineq2} is formally dual to the Generalized Uncertainty Principle, which accounts for ultraviolet quantum gravity corrections, typically by introducing a term proportional to \(\Delta p^2\)~\cite{KMM,Luc1,Luc2}. 
In a similar vein, black hole phenomenology has also been explored within the framework of a generalized Compton wavelength, which arises from a three-dimensional dynamical quantum vacuum~\cite{Pantig:2025bjm}. Collectively, these investigations provide complementary insights into modified gravity theories spanning both the infrared and ultraviolet regimes.

Solving the inequality \eqref{Ineq2} shows that the position uncertainty $\Delta x$ obeys~\cite{Tan}
    \begin{eqnarray}
  &&\frac{L^2\,\Delta p}{\alpha}\left[1-\sqrt{1-\frac{\alpha}{L^2\Delta p^2}}\right]
  \le \Delta x \nonumber\\[2mm]
  &&\hspace{3mm}\le
  \frac{L^2\,\Delta p}{\alpha}\left[1+\sqrt{1-\frac{\alpha}{L^2\Delta p^2}}\right],
\end{eqnarray}
which naturally defines a bounded range within which the position uncertainty \(\Delta x\) must lie. It is straightforward to verify that, for sufficiently small values of the parameter \(\alpha\), the standard uncertainty relation~\eqref{ineq} is consistently recovered.

We now consider Einstein gravity coupled to electromagnetism in AdS spacetime, with an additional quintessence scalar field. The action reads~\cite{Kiselev:2002dx,Sadeghi:2020lfe,Ghosh:2017cuq,Boehmer:2015kta}
\begin{equation}
  S=\int\!d^4x\,\sqrt{-g}\left[\frac{1}{2}(R-\Lambda)-\left(\mathcal{L}_{\rm EM}+\mathcal{L}_{\rm QM}\right)\right],
\end{equation}
where \(R\) is the Ricci scalar, \(\Lambda < 0\) is the cosmological constant associated with AdS spacetime and \(\mathcal{L}_{\rm EM}\) and \(\mathcal{L}_{\rm QM}\) denote the electromagnetic and quintessence Lagrangian densities, respectively, given by
\begin{eqnarray}
  \mathcal{L}_{\rm EM} &=& -\frac{1}{4} F^{\mu\nu} F_{\mu\nu}, \\[2mm]
  \mathcal{L}_{\rm QM} &=& -\frac{1}{2} g^{\mu\nu} \partial_{\mu} \phi\, \partial_{\nu} \phi - V(\phi),
\end{eqnarray}
with $F_{\mu\nu}=\partial_\mu A_\nu-\partial_\nu A_\mu$ is the Faraday tensor of electromagnetic
field.

Varying the total action with respect to the metric yields the field equations
\begin{equation}
  G_{\mu\nu} + \Lambda g_{\mu\nu} = 2\left(T_{\mu\nu}^{\rm EM} + T_{\mu\nu}^{\rm QM}\right),
\end{equation}
where \(T_{\mu\nu}^{\rm EM}\) and \(T_{\mu\nu}^{\rm QM}\) denote the energy-momentum tensors of the electromagnetic and quintessence fields, respectively, and are given by
\begin{eqnarray}
    &T_{\mu\nu}^{\rm EM}&=F_{\mu\lambda}F_{\nu}{}^{\!\lambda}-\frac{1}{4}g_{\mu\nu}F^{2},\\[2mm]
    &T_{\mu\nu}^{\rm QM}&=-\partial_{\mu}\phi\,\partial_{\nu}\phi -g_{\mu\nu}V(\phi).
\end{eqnarray}
Here, $F^2=F^{\mu\nu}F_{\mu\nu}$ (we have set $4\pi G/c^4=1$). 

Solving these equations for a spherically symmetric, charged AdS black hole surrounded by quintessence leads to the following spacetime metric~\cite{Tan}:
\begin{eqnarray}
\label{met}
  ds^2&=&-F(r)\,dt^2 +F(r)^{-1}\,dr^2 +r^2\,d\Omega^2,\\[2mm]
  F(r)&=&1-\frac{2M}{r}+\frac{Q^2}{r^2}+\frac{r^2}{l^2}-\frac{\chi}{r^{3\omega_q+1}},
\end{eqnarray}
where $M$ is the ADM mass, $Q$ is the electric charge, $\chi>0$ the quintessence normalization constant, $d\Omega^2$ the line element on the unit 2-sphere and the barotropic index obeys $-1\le\omega_q\le-\frac{1}{3}$.

In the next section, we apply the above tools to the study of the topological classification of EUP-corrected thermodynamics and photon spheres.

\section{EUP-Corrected RN-AdS Black Hole Thermodynamics with Quintessence}

We consider the thermodynamic modifications of a Reissner–Nordstr\"om–AdS black hole surrounded by quintessence, incorporating corrections from the EUP. The outer event horizon radius \( r_h \) of such system is usually defined as the largest root of the metric function $F(r)$, i.e.,
\begin{equation}
\label{eq:metric_zero}
1-\frac{2M}{r_h}+\frac{Q^2}{r_h^2}+\frac{r_h^2}{l^2}-\frac{\chi}{r_h^{3\omega_q+1}}=0\,.
\end{equation}
Solving for the ADM mass yields
\begin{equation}
\label{ADM}
M(r_h)=\frac{r_h}{2}+\frac{Q^2}{2r_h}+\frac{r_h^3}{2l^2}-\frac{\chi}{2r_h^{3\omega_q}}\,.
\end{equation}

Next, we compute the modifications to thermodynamic quantities, such as temperature and entropy, arising from the EUP. To this end, we employ Hawking's definition of black hole temperature, given by \cite{Xiang:2009yq}
\begin{equation}
T=\frac{\kappa}{8\pi}\frac{\mathrm{d}A}{\mathrm{d}S}\,,
\end{equation}
where the surface gravity is~\cite{Tan}
\begin{equation}
\kappa=\frac{1}{2r_h}\left(1-\frac{Q^2}{r_h^2}+\frac{3r_h^2}{l^2}+\frac{3\omega_q\chi}{r_h^{3\omega_q+1}}\right),
\end{equation}
while, following the argument presented in~\cite{Vagenas}, the Jacobian relating the black hole area to entropy, as modified by the EUP, takes the form
\begin{equation}
\frac{\mathrm{d}A}{\mathrm{d}S}\simeq
16r_h\,\Delta p\,,\,\,\,\,\,\,\,\,\mathrm{with}\,\,\,\Delta p=\frac{1}{4r_h}+\frac{\alpha\,r_h}{L^2}.  
\end{equation}
Hence the EUP‑modified Hawking temperature reads
\begin{equation}
\label{eq:Teup}
T_{\rm EUP}=\frac{r_h}{\pi}\left(\frac{1}{4r_h^2}+\frac{\alpha}{L^2}\right)\left(1-\frac{Q^2}{r_h^2}+\frac{3r_h^2}{l^2}+\frac{3\omega_q\chi}{r_h^{3\omega_q+1}}\right)\,,
\end{equation}
which consistently reduces to the standard expression in the limit of vanishing $\alpha$.

Having assumed \(\alpha > 0\), and in order to ensure that the absolute temperature remains non-negative, we must also require the metric factor to be non-negative. This imposes additional constraints on \(r_h\), which depend explicitly on the value of \(\omega_q\). In particular, for $\omega_q=-1/3$ and $-2/3$ one finds
\begin{eqnarray}
&1-\dfrac{Q^2}{r_h^2}+\dfrac{3r_h^2}{l^2}-\chi\ge0,\qquad(\omega_q=-1/3)\,,\\[2mm]
&1-\dfrac{Q^2}{r_h^2}+\dfrac{3r_h^2}{l^2}-2\chi\,r_h\ge0,\quad(\omega_q=-2/3)\,,
\end{eqnarray}
respectively. The implications of these conditions on black hole dynamics have been explored in~\cite{Tan}, showing that uncertainty principle corrections can lead to incomplete black hole evaporation (see also~\cite{Scar1,Scar2} for alternative approaches that lead to the same conclusion).

Using the expressions~\eqref{ADM} and~\eqref{eq:Teup}, it is now possible to derive the black hole entropy from the first law of thermodynamics, $\mathrm{d}M=T\mathrm{d}S$, obtaining
\begin{eqnarray}
\label{eq:entropy_int}
S&=&\int\frac{\mathrm{d}M}{T_{\rm EUP}}\\[2mm]
\nonumber
&=&\int\frac{(1-Q^2/r_h^2+3r_h^2/l^2+3\omega_q\chi/r_h^{3\omega_q+1})}{2T_{\rm EUP}}\,\mathrm{d}r_h\,,
\end{eqnarray}
which integrates (via a Taylor expansion in small $\alpha$) to
\begin{equation}
\label{eq:Seup}
S_{\rm EUP}=\pi r_h^2\left(1-\frac{2\alpha\,r_h^2}{L^2}\right)+\mathcal{O}(\alpha^2)\,,
\end{equation}
where the integration constant has been fixed so as to properly recover the Bekenstein-Hawking horizon entropy $S=\pi r_h^2$ in the limit \(\alpha \rightarrow 0\).

\section{Thermodynamic topology}
\label{topology}

Recent advances in black hole thermodynamics have highlighted the effectiveness of thermodynamic topology in uncovering and classifying intricate phase structures. Building upon topological methods initially developed by Duan in the context of relativistic particle systems, this framework interprets black holes as thermodynamic topological defects. Within this approach, critical points associated with phase transitions are identified as the zero points of an appropriately defined vector field in the thermodynamic parameter space. These zero points are endowed with integer-valued winding numbers, which serve to characterize the local and global topological properties of the system. This topological classification not only complements traditional thermodynamic analysis but also provides deeper geometric insight into the structure and stability of black hole phase transitions~\cite{Duan:1998jx,Wei:2022dzw,Wei:2023env}.

In this section, we analyze the thermodynamic topology of a charged Reissner-Nordstr\"om AdS black hole surrounded by a quintessential field, incorporating corrections to the black hole entropy arising from the EUP. To carry out this analysis, we adopt the off-shell free energy formalism, defined by
\begin{equation}
\mathcal{F} = M - \frac{S_{\rm EUP}}{\tau}, \label{eq:free_energy}
\end{equation}
where $\mathcal{F}$ denotes the off-shell free energy,  $M$ is the black hole mass, $S_{\rm EUP}$ the EUP-corrected entropy and $\tau$ an inverse temperature parameter defining the thermodynamic ensemble. 

At this point, it is worth emphasizing that thermodynamic topology, enriched by EUP corrections and the effects of the quintessence field, provides a powerful framework for classifying black hole phase transitions. The winding number associated with each critical point directly reflects the stability of the corresponding phase. Specifically,
\begin{itemize}
    \item \( w > 0 \): stable phase (local minimum of \( F \));
    \item \( w < 0 \): unstable phase (local maximum or saddle point of \( F \)).
\end{itemize}
The sum of the winding numbers yields a global topological invariant of the thermodynamic system. As we shall demonstrate, IR corrections shift the locations and properties of the critical points.

From Eq.~\eqref{eq:free_energy}, we can define the two-component vector field
\begin{equation}
\boldsymbol{\phi} = (\phi^{r_h},\,\phi^\Theta) = \left(\frac{\partial \mathcal{F}}{\partial S},\,-\cot\Theta\,\csc\Theta\right), \label{phi vector}
\end{equation}
with coordinates ($r_h, \Theta$) in the extended parameter space. We notice that the zero point of \( \boldsymbol{\phi}^{r_h}\) occurs at $\tau=1/T$,
where \( T \) is the equilibrium Hawking temperature of the black hole in a heat bath.

Let us now construct the unit vector \( n^a = {\phi^a}/{\|\phi\|} \), which satisfies the normalization condition \( n^a n_a = 1 \). We then define the conserved topological current in the three-dimensional parameter space \( (t, S, \Theta) \) as
\begin{equation}
j^\mu = \frac{1}{2\pi}\,\epsilon^{\mu\nu\rho}\,\epsilon_{ab}\,\partial_\nu n^a\,\partial_\rho n^b, \label{eq:topo_current}
\end{equation}
where \( \epsilon^{\mu\nu\rho} \) and \( \epsilon_{ab} \) are the Levi-Civita symbols in 3D and 2D, respectively. The projection of this current onto the \( (\tau,\Theta) \) plane gives the topological density
\[
j^0 = \delta^{(2)}(\boldsymbol{\phi})\,J^0\left(\frac{\boldsymbol{\phi}}{x}\right),
\]
where \( J^0(\boldsymbol{\phi}/x) \) is the Jacobian determinant of the vector field \( \boldsymbol{\phi} \). 

We integrate \( j^0 \) over a compact domain \( \Sigma \) in the $(S,\Theta)$ plane (i.e. entropy-angle space), bounded by smooth contours parametrized as
\begin{equation}
S = S_1 \cos\nu + S_0, \quad \Theta = S_2 \sin\nu + \frac{\pi}{2}, \quad \nu \in (0, 2\pi), \label{eq:contour}
\end{equation}
where $S_1$ ($S_2$) is the amplitude of the oscillation of $S$ ($\Theta$) around the center point $S_0$ ($\pi/2$). In other words, $S_1$ ($S_2$) controls the width (height) of the contour in the entropy (angular) direction.
Although these parameters do not possess fixed or universal numerical values, their choice is guided by functional considerations inherent to the topological analysis. Specifically, \( S_1 \) and \( S_2 \) are selected to be sufficiently small such that the resulting closed contour, as defined in Eq.~\eqref{eq:contour}, encloses only a single zero of the vector field \( {\phi} \). This ensures the accurate isolation of individual critical points and the validity of the local winding number computation. Importantly, the chosen amplitudes must not cause the contour to overlap with multiple singularities, as this would compromise the interpretation of the associated topological charge. Geometrically, \( S_1 \) and \( S_2 \) determine the extension of the integration loop in the entropy and angular directions, respectively, and must be tuned in accordance with the local behavior of the free energy landscape to preserve the locality of the topological classification~\cite{Duan:1998jx,Wei:2022dzw,Wei:2023env}.

The integration of \( j^0 \) over this domain yields the total topological charge
\begin{equation}
W = \int_\Sigma j^0\, dS\, d\Theta = \sum_i w_i, \label{eq:total_charge}
\end{equation}
where each winding number is given by
\begin{equation}
w_i = \frac{1}{2\pi} \oint_{C_i} \epsilon^{ab} n^a\, dn^b, \label{eq:winding}
\end{equation}
and characterizes the net circulation of the vector field around a singular point (defect). These winding numbers are directly associated with the underlying thermodynamic phase.

Combining the definition~\eqref{eq:free_energy} with the expressions in Eqs.~\eqref{ADM} and~\eqref{eq:Seup}, we obtain
\begin{equation}
    \mathcal{F}\hspace{-0.3mm}=\hspace{-0.3mm}\frac{1}{6}\hspace{-0.7mm}\left[\frac{3Q^2}{r_h}+3r_h+8\pi P r_h^3-3\chi r_h^{-3\omega_q}-\frac{6\pi r_h^2}{\tau}\hspace{-0.6mm}\left(1-\frac{2\alpha r_h^2}{L^2}\right)\hspace{-0.6mm}\right]\hspace{-0.4mm},
\end{equation}
with the inverse pressure \( l \) defined as
\begin{equation}
l\equiv\frac{1}{2} \sqrt{\frac{3}{2 \pi P}}\,.
\end{equation}
Therefore, using Eq.~\eqref{phi vector}, the components of the vector $\phi$ can be determined as
	\begin{eqnarray}
   \phi^{r_h}&=&\frac{16\pi\alpha r_h^2-L^2 g(r_h,\tau)}{4\pi\tau\left(L^2-4\alpha r_h^2\right)}\,,\\[2mm]
   \phi^{\Theta}&=&-\cot \Theta\,\csc \Theta\,,
  \end{eqnarray}
  where, in the first relation, we have defined
  \begin{equation}
      g(r_h,\tau)\equiv 4\pi\left(1-2P r_h\tau\right)+\frac{\tau\left[Q^2-r_h\left(r_h+3r_h^{-3\omega_q}\omega_q\chi\right)\right]}{r_h^3} \,.
  \end{equation}
        
We can now identify the zero points of the vector field. One such zero point consistently occurs at \(\Theta = \frac{\pi}{2}\), due to the parameterization~\eqref{eq:contour} of the \(\Theta\)-component of the field. Additionally, 
we derive an equation for \(\tau\) by solving \(\phi^{r_h} = 0\), which gives
\begin{equation}
\tau(r_h)=\frac{4 \pi  r_h^{3 \omega_q +3} \left(L^2-4 \alpha  r_h^2\right)}{L^2 \left[r_h^{3 \omega_q } \left(8 \pi  P r_h^4-Q^2+r_h^2\right)+3 r_h \chi  \omega_q \right]},
\label{taurh}
\end{equation}
In examining the behavior of \( \tau(r_h) \), we identify its inflection points, which manifest critical behavior typically associated with phase transitions or qualitative changes in the thermodynamic structure of the system~\cite{Wei:2023env}. These points satisfy the condition $\tau''(r_h)=0$
and indicate a change in the concavity of \( \tau(r_h) \). 

For the borderline quintessence value $\omega_q=-\tfrac13$ (corresponding to \(3\omega_q + 3 = 2\)), the function \eqref{taurh} takes the form
\begin{equation}
    \tau_{\omega_q=-1/3}(r_h)
=\frac{4\pi\,r_h^2\,(L^2-4\alpha\,r_h^2)}
{L^2\bigl[r_h^{-1}(8\pi P\,r_h^4 -Q^2 +r_h^2)-r_h\,\chi\bigr]}\,.
\end{equation}
By assigning characteristic values to the parameters, specifically \( L = 1 \), \( P = 0.1 \), \( Q = 0.1 \), \( \chi = 1 \), and \( \alpha \sim \mathcal{O}(10^{-1}) \), it can be shown that $\tau''(r_h) \neq 0 \quad \forall\, r_h$,
and that the equation \( \tau'(r_h) = 0 \) has no real solutions. Therefore, no inflection points or phase-transition structures occur.

The thermodynamic topology for the case \( \omega_q = -1/3 \) is shown in detail in Fig.~\ref{b}. In particular, panel~\ref{fig2a} displays the corresponding monotonic curve in the \((r_h, \tau)\) plane, within the off-shell free energy ensemble. This curve represents what we refer to as the ``defect curve''. Since there is no point at which \( \tau''(r_h) \) vanishes, no thermodynamic critical radius emerges in this context, as previously discussed.
Figure~\ref{fig2b} illustrates the normalized vector field \( n^a = \bigl( \partial_{r_h} \tau, -\partial_{\Theta} \tau \bigr) / |\varphi| \), defined over the domain \( (r_h, \Theta) \in [r_{\min}, r_{\max}] \times [0, \pi] \).
Since \( \tau \) is independent of \( \Theta \), we have \( \partial_{\Theta} \tau = 0 \) throughout the entire domain. Moreover, as \( \partial_{r_h} \tau > 0 \) everywhere in the interior, the vector field \( \varphi^a \) has no interior zeros. The only point where \( |\varphi| \) vanishes lies on the boundary at \( \Theta = \pi/2 \), giving rise to a single, unbroken contour.
In Fig.~\ref{fig2c}, we display the phase angle \( \Omega = \arg\bigl( \varphi^{r_h} + i\, \varphi^\Theta \bigr) \) around this single limiting zero. Traversing a small counterclockwise loop around this point results in an increase of \( \Omega \) by \( +2\pi \), corresponding to a winding number $w=(2\pi-0)/2\pi=1$ and a topological charge $W = +1$.

\begin{figure}[H]
\begin{center}
\subfigure[~Plot of $\tau$ versus $r_h$. We set $\alpha=0.4, P=0.1, Q=0.1, L=1$ and $\chi=1$.]  {\label{fig2a}
\centering\includegraphics[width=0.4\textwidth]{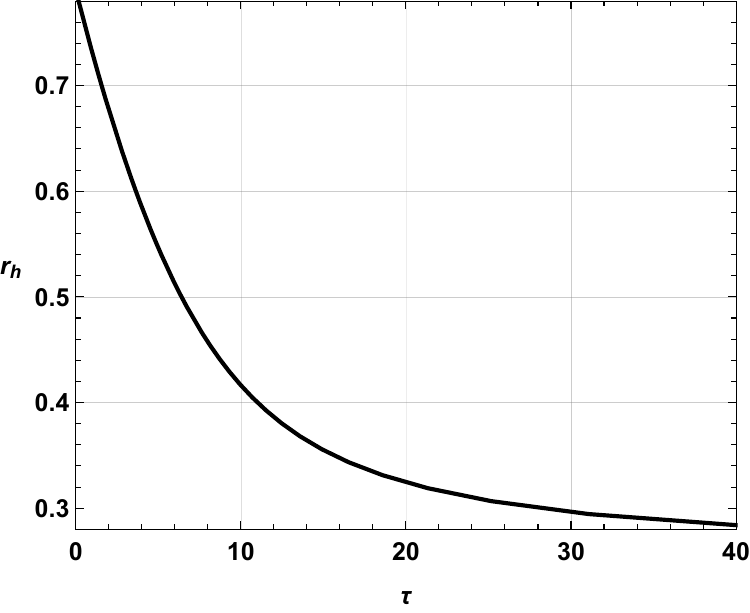}}\hfill
\subfigure[~The vector plots of the normalized vector
field, with the zero points located at $\theta = \pi/2$ under the EUP entropy.]  {\label{fig2b}
\centering\includegraphics[width=0.4\textwidth]{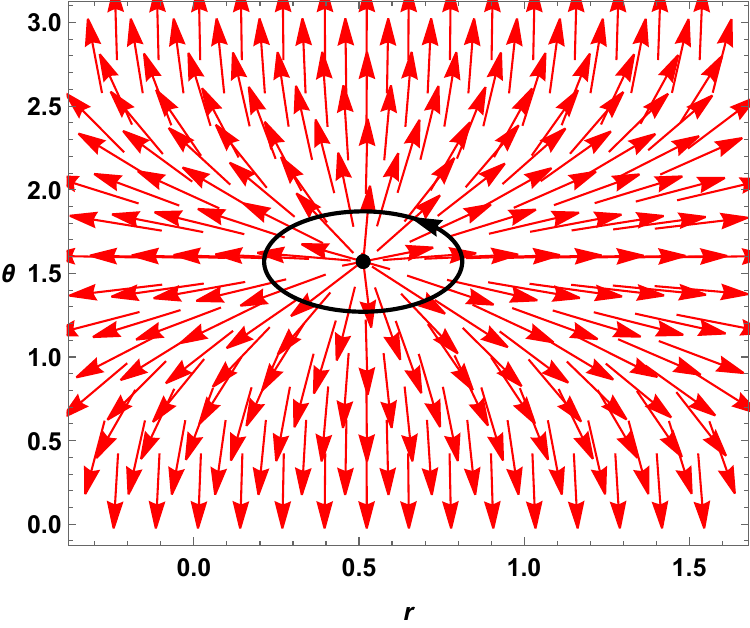}}
\end{center}
\begin{center}
\subfigure[~Phase winding $\Omega(\theta)$.]  {\label{fig2c}
\centering\includegraphics[width=0.47\textwidth]{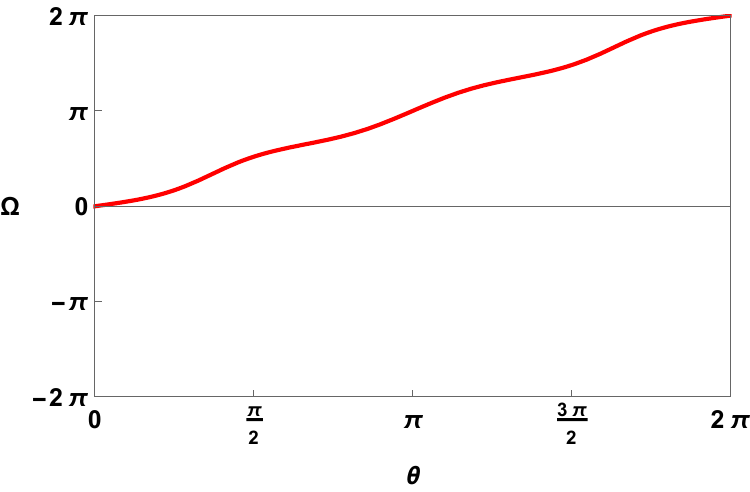}}
\end{center}\label{c}
\caption{Thermodynamic topology for $\omega_q=-1/3$.}.\label{b}
\end{figure}

On the other hand, for the canonical quintessence case \(\omega_q=-\frac{2}{3}\), the exponent \(3\omega_q+3\) reduces to unity. In this case, Eq. \eqref{taurh} becomes
\begin{equation}
    \tau_{\omega_q=-2/3}(r_h)
=\frac{4\pi\,r_h\,(L^2-4\alpha\,r_h^2)}
{L^2\bigl[r_h^{-2}(8\pi P\,r_h^4 -Q^2 +r_h^2)-2\,r_h\,\chi\bigr]}\,.
\end{equation}

Under the same parameter assumptions as before, one finds that both the first and second derivatives of the resulting rational function exhibit two distinct positive zeros. This behavior signals the emergence of an intermediate, quasi-stable branch in the solution space, induced by the EUP corrections. Specifically, the inflection points identified by the second derivative delineate the boundaries between the small-, intermediate-, and large-black hole branches, indicating the presence of a first-order phase transition driven by the EUP parameter.

Interestingly, increasing the value of \( \alpha \) causes both inflection points to move closer together, thereby narrowing the intermediate black hole branch. In contrast, for \( \omega_q = -\tfrac{1}{3} \), variations in \( \alpha \) do not lead to any qualitative changes, as no extrema or inflection points emerge in this case, and the phase structure remains trivial.
}

Physically, shifts in the inflection radii correspond to percent-level modifications in the quasinormal mode frequencies and the photon sphere radii. In particular, for \( \omega_q = -\frac{2}{3} \), the separation of light ring radii induced by the EUP leads to the formation of two distinct photon spheres—one stable and one unstable—characterized by winding numbers \( \pm 1 \) (see below for further discussion).  Notably, the resulting shifts in the shadow diameter, amounting to a few percent for supermassive black holes such as M87*, lie within the sensitivity range of current Event Horizon Telescope (EHT) observations~\cite{M87}. In this context, the EHT image of the M87* shadow has recently been employed to test the validity of fundamental physics in the strong-field regime, particularly highlighting potential violations of the No-Hair Theorem~\cite{Khodadi:2021gbc}.

\begin{figure}[t]
\begin{center}
{\label{Metric_Function_RN_AdSa_4}
\centering\includegraphics[width=0.4\textwidth]{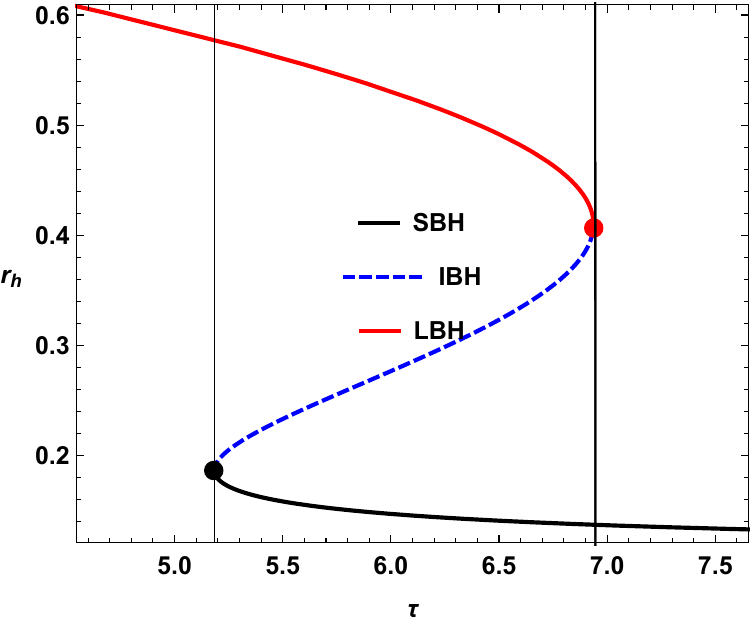}}
\end{center}
\caption{Defect curve of quintessential charged AdS black holes with EUP corrected entropy for $\omega_q=-2/3$. We set $\alpha=0.4, P=0.1, Q=0.1, L=1$ and $\chi=1$.}\label{4}
\end{figure}

To gain deeper insight into the case $\omega_q = -2/3$, we examine the behavior of $\tau(r_h)$, which - as illustrated in Fig.~\ref{4} - exhibits three different branches. This suggests a more intricate two-phase structure involving small, intermediate (unstable) and large black hole phases. The existence of multiple turning points signals the onset of metastable behavior.

\begin{figure}[H]
\begin{center}
\subfigure[~1$^{st}$ contour: C$_1$.]  {\label{Metric_Function_RN_AdSa_7}
\centering\includegraphics[width=0.4\textwidth]{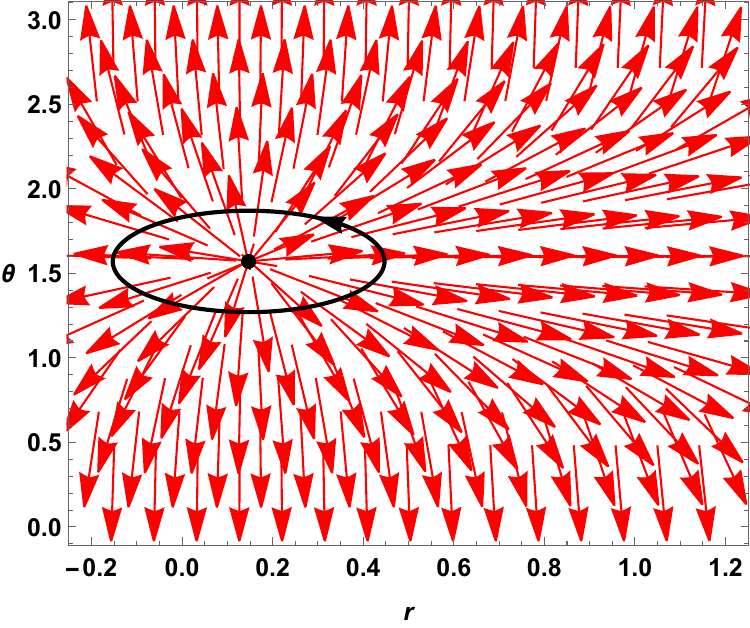}}\hfill
\subfigure[~2$^{th}$ contour: C$_2$.]  {\label{MetricFunctionRNAdSa8}
\centering\includegraphics[width=0.4\textwidth]{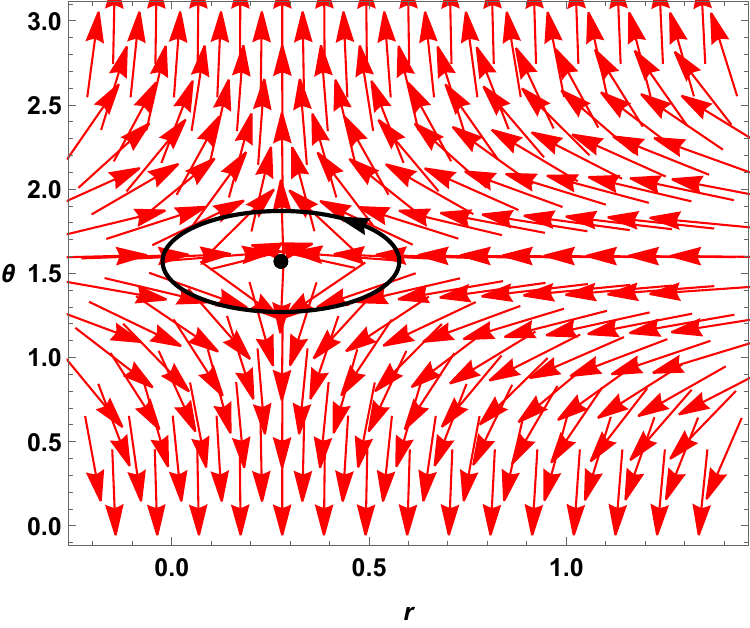}}\hfill
\subfigure[~3$^{th}$ contour: C$_3$.]  {\label{Metric_Function_RN_AdSa_9}
\centering\includegraphics[width=0.4\textwidth]{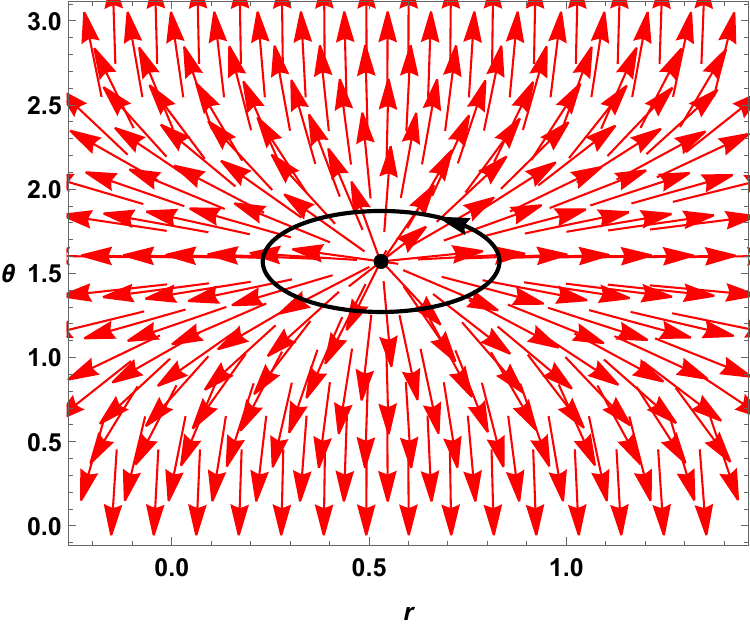}}\hfill
\end{center}
\caption{The vector plots of the normalized vector
field, with the zero points located at $\theta = \pi/2$ under the EUP entropy ($\omega_q=-2/3)$.}\label{5}
\end{figure}

This feature can also be confirmed from an alternative perspective by analyzing the vector field $n^a(\tau, \Theta)$, as shown in Fig.~\ref{5}.
Specifically, along the line $\Theta = \pi/2$, two zero points emerge, giving rise to three distinct closed contours. The outer zeros, labeled $C_1$ and $C_3$, correspond to winding numbers $w = +1$ and $w = -1$, respectively. In contrast, the central contour, denoted as $C_2$, is associated with a vanishing winding number, $w = 0$, since it does not enclose any critical point. Altogether, this configuration yields a richer topological structure characterized by a total topological charge of $W = 0$.

To better understand the above features, Fig.~\ref{5} presents a zoomed-in view of the three contours for $\omega_q = -2/3$, with particular emphasis on the behavior of the vector field. The contour $C_2$ clearly exhibits a zero-winding behavior, characterized by an inward spiral structure, while the outer contours, $C_1$ and $C_3$, display single-defect behaviors associated with winding numbers $w_1 = +1$ and $w_3 = -1$, respectively. This analysis confirms the winding number assignments: $w_1 = +1$, $w_2 = 0$, and $w_3 = -1$. 

\begin{figure}[H]
\begin{center}
\subfigure[]  {\label{Metric_Function_RN_AdSb_10}
\centering\includegraphics[width=0.45\textwidth]{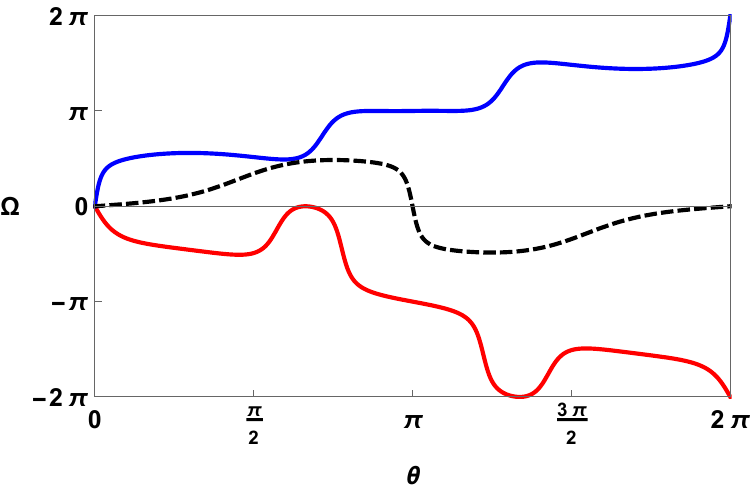}}
\end{center}
\caption{Phase winding $\Omega(\theta)$ for $\omega_q=-2/3$. The blue, black and red curves correspond to the contours $C_1$, $C_2$ and $C_3$, respectively.}\label{6}
\end{figure}

The winding number integrals are presented in Fig.~\ref{6}. A direct computation of the corresponding line integrals further verifies the winding numbers around the critical points. In particular, the winding numbers associated with the three contours can be explicitly computed from the total change in the angle $\theta$ along each path. For the contour $C_1$ (blue curve), the total variation of the angle is $2\pi$, leading to a winding number $w_1 = (2\pi - 0)/(2\pi) = 1$. For the contour $C_2$ (black curve), the angle does not change, yielding $w_2 = (0 - 0)/(2\pi) = 0$. Finally, for the contour $C_3$ (red curve), the angle decreases by $2\pi$, resulting in a winding number $w_3 = (-2\pi - 0)/(2\pi) = -1$.

\section{Photon Spheres}
A photon sphere is a region consisting of null geodesics that forms the boundary for any stable photon orbits within the strong gravitational field of a black hole. It represents the location where light undergoes extreme bending due to the intense curvature of spacetime. 

Photon spheres can manifest in two distinct forms: stable and unstable. In the unstable case, small perturbations can cause photons to either escape to infinity or fall into the black hole. This unstable photon sphere plays a fundamental role in the study of black hole shadows, as it delineates the critical boundary between light that escapes and light that is captured by the black hole. On the other hand, a stable photon sphere resists perturbations, allowing photons to remain confined to this orbit. This feature makes it particularly relevant for investigating spacetime instabilities. 

Traditionally, the analysis of photon spheres involves deriving the Lagrangian from the action, which is subsequently used to construct the Hamiltonian governing the system’s dynamics. The Hamiltonian plays a crucial role in defining the effective potential, which depends on the particle's energy and angular momentum. This potential is fundamental in determining the properties of the photon sphere.

In this work, we adopt a geometrically motivated approach based on the effective potential introduced by Cunha et al.~\cite{P1} for spherically symmetric ultra-compact objects. This potential has been extensively employed to investigate photon spheres in four-dimensional black holes, including those in asymptotically AdS and dS spacetimes. For a detailed discussion of its derivation and applications, we refer the reader to Refs.~\cite{P1,Wei:2020ght,P3,P4,P5}. 

As emphasized in the Introduction, the study of the photon sphere primarily concerns the local geometrical properties of the black hole spacetime, particularly in the vicinity of the unstable circular null orbits. In this regime, the dominant contributions arise from the immediate metric structure around the photon sphere radius, where quintessence-induced modifications can be significant due to their effect on the effective potential for photons. On the other hand, we expect that the EUP-induced corrections, which are inherently infrared in nature, have a negligible impact at these relatively small  length scales. Therefore, in the following analysis, it is justifiable to neglect such corrections while retaining quintessence effects to capture the relevant physics \cite{Peng:2020wun}.

A key feature of the effective potential introduced by Cunha
is that it depends solely on the spacetime geometry, making it independent of the particle’s conserved quantities such as energy and angular momentum. This property enables a more direct and intrinsic analysis of the gravitational field surrounding ultra-compact objects. By projecting the associated vector field ${\phi}$ - which captures the spatial variations of the potential - onto the equatorial plane and exploiting the spacetime’s spherical symmetry, the problem effectively reduces to a lower-dimensional analysis\footnote{Although the vector field associated with the photon sphere is denoted by \( v \) in the original analysis~\cite{Cunha2}, we follow the notation of Ref.~\cite{Wei:2020ght} and label it as \( \phi \) to align with Duan's topological current \( \phi \)-mapping theory.}. This simplification facilitates a systematic classification of different regions according to the geometric properties of the photon sphere.

In this framework, photon spheres correspond to the critical points of the effective potential. The associated topological charge can be computed through an integral involving a Dirac delta function, which localizes the contribution precisely at the zeros of the vector field. Thanks to the spherical symmetry, the analysis reduces to studying the radial component, whose zeros directly identify the locations of photon spheres, each characterized by a well-defined topological charge $Q$.

In the following, we apply this framework to investigate the structure and properties of photon spheres and light rings around charged AdS black holes surrounded by a quintessence field. We begin by introducing the essential definitions and fundamental equations required for this analysis. 
To this end, we follow \cite{Wei:2020ght} and consider the black hole solution in the following general form:
\begin{equation}
ds^2 = -f(r) \, dt^2 + \frac{1}{g(r)} \, dr^2 + h(r) \left( d\theta^2 + \sin^2 \theta \, d\varphi^2 \right) \,.
\end{equation}
This metric will later be specialized to the specific case given in Eq.~\eqref{met}. 

Focusing on the motion of light (null geodesics), we restrict our attention to the radial equation of motion in the equatorial plane~\cite{P1,Wei:2020ght,P3,P4,P5}, which takes the form
\begin{equation}\label{Ps1}
\dot{r}^2 + V_{\text{eff}}(r)=0\,,
\end{equation}
where the effective potential \( V_{\text{eff}} \) is given by
\begin{equation}\label{Ps2}
V_{\text{eff}} = g(r) \left[ \frac{L^2}{h(r)} - \frac{E^2}{f(r)} \right].
\end{equation}
Here, $E$ and $L$ denote the photon's energy and angular momentum, associated with the Killing vectors $\partial_t$ and $\partial_\phi$, respectively. 

Due to spherical symmetry of the solution, there exists a radius $r_{ps}$ corresponding to the photon sphere, determined by the conditions \cite{P1,Wei:2020ght,P3,P4,P5}
\begin{equation}\label{Ps3}
V_{\text{eff}} = 0, \quad \partial_r V_{\text{eff}} = 0,
\end{equation}
which lead to
\begin{equation}\label{Ps4}
\left( \frac{f(r)}{h(r)} \right)' \bigg|_{r = r_{ps}} = 0.
\end{equation}
Here, the prime denotes differentiation with respect to $r$. 
The stability of the photon sphere is assessed via the second derivative: $\partial_r^2 V_{\text{eff}}(r_{ps}) < 0$ indicates instability, whereas $\partial_r^2 V_{\text{eff}}(r_{ps}) > 0$ implies stability. 

Differentiating Eq. \eqref{Ps4}, we obtain
\begin{equation}\label{Ps5}
f(r) h'(r) - f'(r) h(r) = 0\,.
\end{equation}
At the black hole horizon \( r = r_h \), we have \( f(r_h) = 0 \), which causes the first term to vanish. However, since \( f'(r_h) \) is generally nonzero, it follows from Eq.~\eqref{Ps5} that \( r_{ps} \neq r_h \). An exception occurs in the extremal black hole case, where the horizon becomes degenerate, i.e., $f(r_h) = 0$ and $f'(r_h) = 0$, causing the photon sphere to coincide with the horizon. 

To investigate the topology of the photon sphere, we introduce a potential function that is regular throughout the spacetime, i.e. \cite{P1}, 
\begin{equation}
\label{Ps6}
H(r, \theta) = \frac{1}{\sin \theta} \sqrt{\frac{f(r)}{h(r)}}\,.
\end{equation}
Clearly, the condition $\partial_r H = 0$ identifies the photon sphere radius.

Following the approach outlined in Ref.~\cite{Cunha2}, we define a vector field \( {\phi} = (\phi^r, \phi^\theta) \) in the form
\begin{equation}\label{Ps7}
\phi^{r} =  \sqrt{g(r)} \, \partial_r H, \quad \phi^\theta = \frac{\partial_\theta H}{\sqrt{h(r)}}\,.
\end{equation}
For spherically symmetric spacetimes, the circular photon orbit is independent of the angular coordinate $\theta$, though we retain $\theta$ for completeness in studying the topology. This vector field can also be expressed in complex form,
\begin{equation}\label{Ps8}
\phi = \phi^{r} + i \phi^\theta = \|\phi\| e^{i \Theta}\,.
\end{equation}
At the photon sphere, where \( {\phi} = 0 \), the phase \( \Theta \) is not well-defined. Therefore, we consider \( {\phi} \) directly as a complex vector field.
The normalized components are
\begin{equation}\label{Ps9}
n^a = \frac{\phi^a}{\|\phi\|}, \quad a=1,2\,,
\end{equation}
where $||\phi||=\sqrt{\phi_a\phi^a}$ and $\phi^1 = \phi^{r}$, $\phi^2 = \phi^\theta$.

Using the above construction, we now investigate the behavior of photon spheres. A crucial topological property is that a zero point of ${\phi}$ within a closed loop contributes a topological charge equal to the winding number. Each photon sphere is assigned a charge of either $+1$ or $-1$, depending on the winding direction. Depending on the chosen loop (which may encircle one or multiple zero points), the total topological charge can be $-1$, $0$, or $+1$. In classical black hole scenarios where $M > Q$, the expected structure of photon sphere yields a total topological charge of $-1$. This configuration is consistent with the Weak Cosmic Censorship Conjecture and ensures that a proper event horizon and photon sphere do exist.  

\begin{figure}[H]
 \begin{center}
 \subfigure[]{
\centering\includegraphics[width=0.35\textwidth]{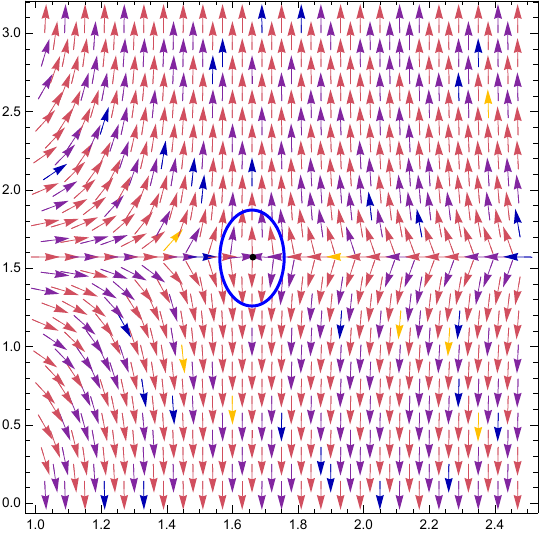}
 \label{100a}}
 \subfigure[]{
\centering\includegraphics[width=0.35\textwidth]{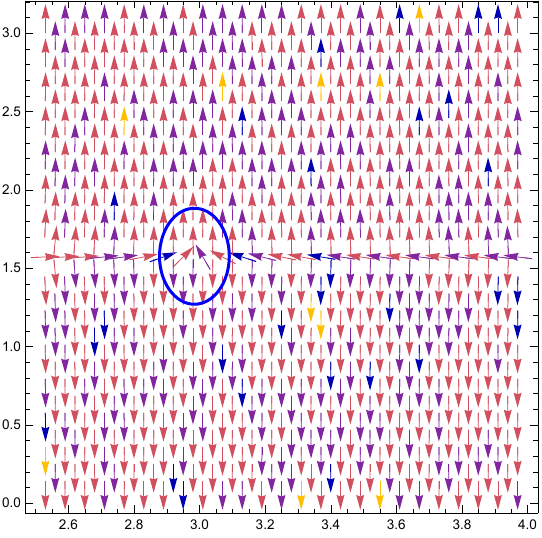}
 \label{100b}}
 \subfigure[]{
\centering\includegraphics[width=0.35\textwidth]{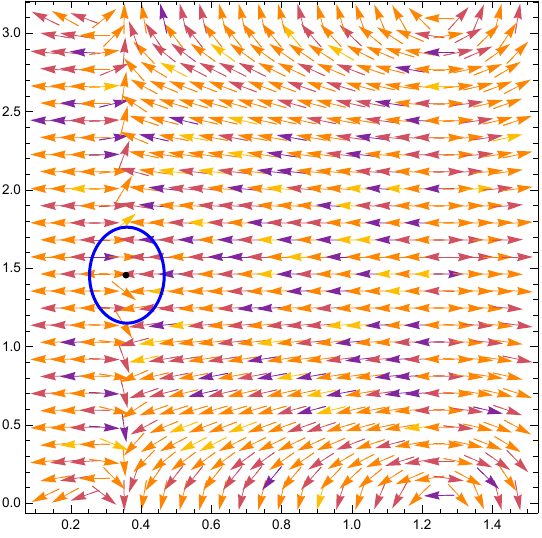}
 \label{100c}}
 \caption{\small{The plot of photon spheres of charged AdS black holes immersed in a quintessence field: Fig. \ref{100a} with respect to ($\chi=0.1, Q=0.1$),  Fig. \ref{100b} with respect to ($\chi=0.5, Q=0.1$) and  Fig. \ref{100c} with respect to ($\chi=1, Q=1$) for $\omega_q=-1/3$ and $l=1$.}}
 \label{m1}
\end{center}
 \end{figure}

\begin{figure}[H]
 \begin{center}
 \subfigure[]{
\centering\includegraphics[width=0.35\textwidth]{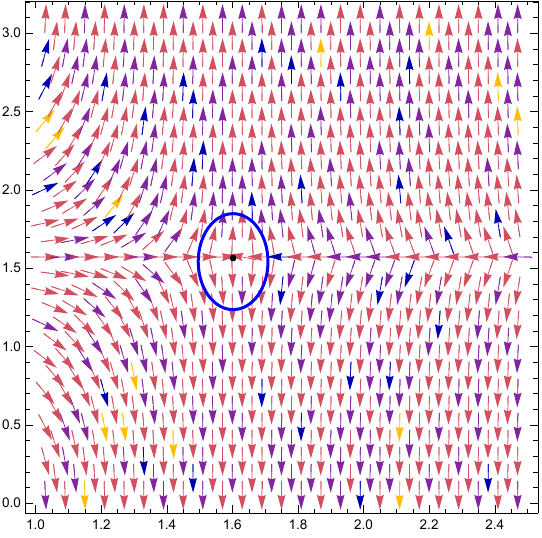}
 \label{200a}}
 \subfigure[]{
\centering\includegraphics[width=0.35\textwidth]{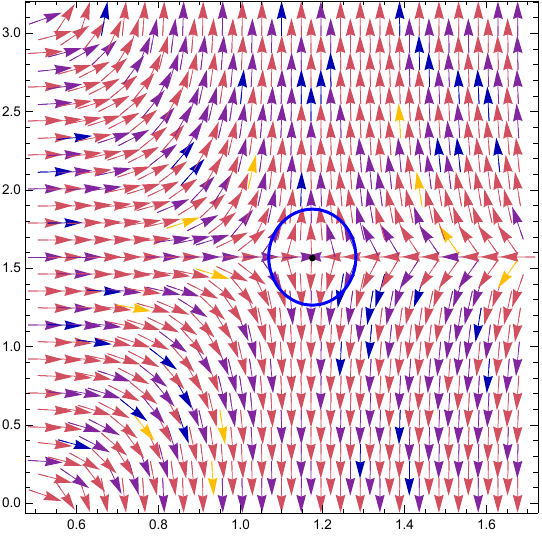}
 \label{200b}}
 \subfigure[]{
\centering\includegraphics[width=0.35\textwidth]{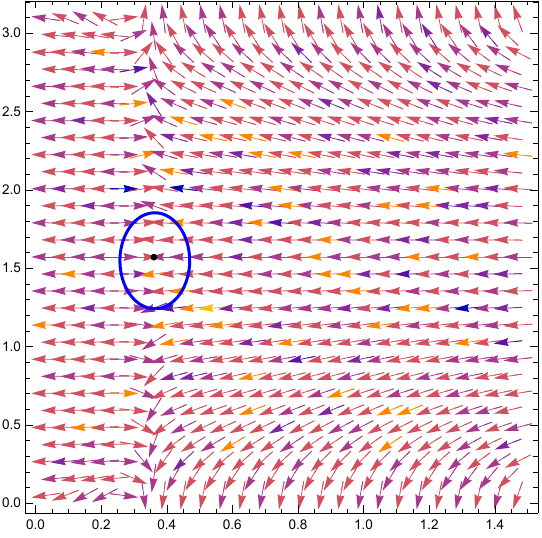}
 \label{200c}}
 \caption{\small{The plot of photon spheres of charged AdS black holes immersed in a quintessence field: Fig. \ref{200a} with respect to ($\chi=0.1, Q=0.1$),  Fig. \ref{200b} with respect to ($\chi=0.5, Q=0.1$) and  Fig. \ref{200c} with respect to ($\chi=1, Q=1$) for $\omega_q=-2/3$ and $l=1$.}}
 \label{m2}
\end{center}
 \end{figure}

With respect to the metric function \eqref{met}, we can calculate the components of the vector field as follows:
\begin{eqnarray}
\nonumber
\phi^{r}&=&\frac{\csc (\theta )}{2r^4}
\left\{r\left[6M-2r+3\chi r^{-3\omega_q}\left(1+\omega_q\right)\right]-4Q^2
\right\},
\\
\label{Ps10}
\\[2mm]
\nonumber
\phi^{\theta }&=&-\frac{\cot (\theta ) \csc (\theta )}{r} \sqrt{\frac{1}{l^2}+\frac{Q^2-r\left(2M-r+\chi r^{-3\omega_q}\right)}{r^4}}\,.
\\
\label{Ps11}
\end{eqnarray}
In order to investigate how this structure varies with different parameters, 
we analyze the behavior of the photon sphere for various values of the free parameters: \( Q = 0.1, 1 \), \( l = 1 \) and \( \chi = 0.1, 0.5, 1 \).
The results, illustrated in the figures below, show that for $\omega_q=-1/3, -2/3$, the system maintains a total photon sphere topological charge of \( -1 \) (see Figs.~\ref{m1} and \ref{m2}). Thus, the investigation of photon spheres deepens our understanding of the geometric and physical properties of the black holes under consideration.

Building on these insights, exploring the relationship between the topological charges of black holes~\cite{Wei:2023env,Wei:2022dzw,20a,21a,22a,24a,26a,27a,28a,33a,34a,35a,37a,38',38c,39a,40a,43a,44f,44g,44h,44i,44j,44k,44l,44m} and the topology of photon spheres presents a promising direction for future research. Such investigations could shed new light on gravitational lensing and black hole shadow formation, with far-reaching implications for astrophysical observations.

\subsection{$\omega=-\frac{1}{3}$}
 \subsection{$\omega=-\frac{2}{3}$}

\section{Conclusions and Outlook}
In this work we have carried out a comprehensive investigation of charged AdS black holes surrounded by quintessence under EUP corrections, focusing on the two representative indices \(\omega_q=-2/3\) and \(\omega_q=-1/3\).  We derived closed-form expressions for the EUP-modified Hawking temperature and entropy, demonstrating that the EUP deformation parameter \(\alpha\) non-trivially delays the onset of thermodynamic instability and shifts critical radii.
The inflection-point analysis of the relaxation-time function \(\tau(r_h)\) revealed two real inflection points for \(\omega_q = -2/3\) (indicative of small, intermediate, and large horizon phases), while no inflection point emerges for \(\omega_q = -1/3\).
Furthermore, by employing Duan's $\phi$-mapping topological current approach, we identified conventional $(+1)$, $(-1)$ and novel $(0)$ winding numbers at each thermodynamic critical point. 

In parallel, we have investigated how the structure of photon spheres varies with different parameter choices, focusing in particular on the behavior for selected values of the  parameters $Q, l$ and $\omega_q$.
Our findings reveal that the system consistently maintains a total photon sphere topological charge of \( -1 \). This result enhances our understanding of the geometric and physical characteristics of the black holes under study, particularly regarding the structure and effective geometry of photon orbits.

In light of the results obtained, it becomes evident that the connection between the topological charges of black holes and the topology of photon spheres offers compelling prospects for future research in gravitational lensing and shadow formation, with profound implications for astrophysical observations. Our detailed illustrations and mathematical derivations further underscore the influence of variations in the model parameters on the thermodynamic topology of black holes. The results consistently suggest that stable black hole configurations correspond to positive winding numbers, whereas negative winding numbers are indicative of instability. These findings not only advance our theoretical understanding of black hole stability but also establish a coherent framework for interpreting such stability through the lens of topological invariants.

Moreover, our analysis illustrates that large-scale quantum-gravity corrections and dark-energy fields can quantitatively reshape black hole phase diagrams and light-ring configurations, producing percent-level shifts in quasinormal-mode frequencies and shadow diameters, while leaving intact the global topological invariants that classify their thermodynamic and geometric structures. This robust topological framework thus provides a promising avenue for confronting EUP phenomenology and quintessence models with forthcoming high-resolution black hole shadow imaging and gravitational-wave observations. Further investigations along these lines are underway and will be presented in future work.

\section*{Acknowledgements}
The research of GGL is supported by the postdoctoral fellowship program of the University of Lleida. GGL gratefully acknowledges the contribution of the LISA Cosmology Working Group (CosWG), as well as support from the COST Actions CA21136 - \textit{Addressing observational tensions in cosmology with systematics and fundamental physics (CosmoVerse)} - CA23130, \textit{Bridging high and low energies in search of quantum gravity (BridgeQG)} and CA21106 -  \textit{COSMIC WISPers in the Dark Universe: Theory, astrophysics and experiments (CosmicWISPers)}.

\end{document}